\documentclass[11pt, nofootinbib]{revtex4}
\usepackage{hyperref}
\usepackage{tikz}
\usepackage{amsmath,amssymb}
\def\be {\begin{equation}}
\def\ee {\end{equation}}
\def\bea {\begin{eqnarray}}
\def\eea {\end{eqnarray}}
\def\bc {\begin{center}}
\def\ec {\end{center}}
\def\bfg {\begin{figure}}
\def\efg {\end{figure}}
\def\bi {\begin{itemize}}
\def\ei {\end{itemize}}
\def\nn {\nonumber}

\def\le {\left}
\def\ri {\right}

\def\beq{\begin{equation}}
\def\eeq{\end{equation}}
\def\br{\begin{eqnarray}}
\def\er{\end{eqnarray}}
\newcommand{\eel}[1] {\label{#1}\end{equation}}

\begin{document}

\title{Hawking Radiation Power Equations for  Black Holes}
\author{Ravi Mistry$^a$}
\email{ravi.mistry.r@gmail.com}
\author{Sudhaker Upadhyay$^b$}
\email{sudhakerupadhyay@gmail.com}
\author{Ahmed Farag Ali$^c$}
\email{ahmed.ali@fsc.bu.edu.eg}
\author{ Mir Faizal$^d$}
 \email{mirfaizalmir@gmail.com}

\affiliation{$^a$Institute of Physics, University of  Brasilia, Brasilia, DF, 70910-900, Brazil}
\affiliation{$^b$Centre for Theoretical Studies, Indian Institute of Technology Kharagpur, Kharagpur-721302, India}
\affiliation{$^c$Netherlands Institute for Advanced Study
Korte Spinhuissteeg 3, 1012 CG Amsterdam, Netherlands and\\
Department  of Physics, Faculty of Science,  
 Benha University, Benha, 13518, Egypt}
\affiliation{$^d$Department of Physics and Astronomy, University of Lethbridge,
Lethbridge, Alberta, T1K 3M4, Canada and\\
Irving K. Barber School of Arts and Sciences, University of British Columbia - Okanagan,   3333 University Way, Kelowna,   British Columbia V1V 1V7, Canada}

\begin{abstract} 
 { We derive the Hawking radiation power equations for black holes in  asymptotically flat,
asymptotically Anti-de Sitter (AdS)  and asymptotically   de Sitter  (dS) black holes, This is done by using the greybody factor for these black holes. 
We observe that   the radiation power equation  for asymptotically flat black holes,  corresponding to greybody factor  at low frequency, depends on both  the  Hawking temperature and the horizon radius. However, for the greybody factors at asymptotic  frequency, it only depends  on the  Hawking temperature. We also obtain the power equation for 
 asymptotically AdS black holes  both below and above the  critical frequency. 
The radiation power equation   for  at asymptotic frequency 
 is same for both Schwarzschild AdS and Reissner-Nordstr\"om AdS solutions and only depends on the   Hawking temperature.
We also discuss the   power equation for asymptotically dS black holes at low frequency,   for both even or odd  dimensions.}
 \end{abstract}
 \maketitle

\section{Introduction}
The  evaporation of a black hole can be understood in terms of the black body factor and the greybody factor. The  black body factor is calculated from   the probability of  a particle being created in the vicinity of a horizon,
and the greybody factor is calculated from the   probability that
this particle penetrates the potential barrier and escapes to
infinity. The analysis of the black hole greybody factors shows that Hawking emission from a highly rotating black hole is strongly spin dependent, with particles of highest spin (gravitons) dominating the energy spectrum  \cite{1}. 
Gravitational greybody factors are analytically computed for static, spherically symmetric
black holes, including black holes with charge and in the presence of a cosmological
constant \cite{20}. In this context, the greybody factors for both asymptotically dS and AdS spacetimes can be obtained.  There are many
distinct models with  exact black hole solutions in literature, which in turn implies the need for concrete calculations of the corresponding greybody
factors (some recent developments can be found in \cite{3,4,5,6,7,8,9,10,11,12,13,14}).

These greybody factors were studied  long ago in \cite{41, 51}. 
one can easily follow the basic  set-up of the calculations.
There  might be some trouble in order to obtain exact results with such set-up. 
The
  scattering problem of black holes has common features of
  scattering in media with some index of refraction. In fact, 
  the curvature of space-time itself
involved in the scattering of black holes. \cite{61}.  
The
quantum Hawking evaporation of near-extremal Reissner-Nordstr\"om   black holes
is studied recently, where  the effective curvature potential  causes distortion 
in   the familiar radiation
spectrum of genuine $(3+1)$-dimensional perfect black-body emitters \cite{71}.
The Hawking radiations of black holes in asymptotically flat, AdS and dS spacetimes through greybody factors  are  subject of interesting investigations.
With the help of radiation power  expression, one can compute total energy emitted as Hawking radiation by multiplying the power with black hole evaporation time scale and this must be equal to the total mass of the black hole by virtue of 
energy conservation. This may provide an  insight to the process of black hole evaporation.

It may be noted that we use a formalism that depends on  the frequency, and can be consistently applied  to the greybody factors which are functions of all frequencies   \cite{20}.  So, we use a form of the greybody factor which is defined for all the frequencies. Here we note that the low energy expression for the greybody factor for scalar fields in the background of a higher-dimensional Schwarzschild black hole have been obtained    derived in \cite{P1}.  The low-energy greybody
factor for a higher-dimensional dS black hole have also been studied  \cite{P2}. The emission of Hawking radiation of higher-dimensional
black holes in the bulk (greybody factors and
radiation spectra) are studied in details for the emission of scalar modes, and the ratio of the
missing energy over the visible one is calculated for different values of the number of extra
dimensions \cite{P3}.
The procedure of scalar wave absorption by a black hole is followed here.
This is because of nature of  scalar wave which  spreads from infinity
over whole spacetime and  becomes reflected by the black hole potential barrier.
The transmitted scalar wave, near the  horizon, appears as the incoming radiation.  In fact, for low frequency scattering, the greybody factor  is found equal to the absorption probability of the black hole. The reason to consider the real frequencies  at low
energy is the opposite nature of scattering and absorption processes.   
Our work is   mainly based on the results of \cite{20}, where it is emphasized that the main contribution to the greybody factor of black holes comes from the $l = 0$ mode   in the low frequency limit.

With this motivation, in this paper,   we calculate the Hawking radiation power equation for black holes in $d+1$-dimensional asymptotically flat, AdS and dS spacetimes  with the help of  following equation \footnote{We use the unit system in which $G=\hbar=c=k_{B}=1$.} \cite{zur,page}
\bea
P^{d+1}=\frac{T_H}{2\pi}\int ^{\infty}_0 d\omega\frac{\gamma(\omega)x}{e^{x}-1},\label{HRPE}
\eea 
where $x\equiv\frac{\omega}{T_H}$, $T_H$ is the Hawking temperature and $\gamma(\omega)$~ {represents greybody factor}.   
{First, we consider   the greybody factor at the low frequency limit  for  black holes in asymptotically flat spacetime and compute the respective Hawking radiation power equation which depends on both, the  Hawking temperature and   horizon radius $R_H$ under different 
power law. Then, we derive the Hawking radiation power equation corresponding to  the greybody factor  at asymptotic frequency for Schwarzschild solution  and find that it depends on the  Hawking temperature only and does not depend on horizon radius. Interestingly, we observe that
the radiation power equations corresponding to greybody factors at asymptotic frequency  are same for all spacetime dimensions as it does not depend on the dimension $d$. In fact,  it depends on
Hawking temperature only with power law.  Further, we compute the Hawking radiation power equation along greybody factor at the low frequency limit  for asymptotically AdS black holes. We consider here both cases for which the frequencies are much lower and higher than the critical frequency (at which the black hole absorbs all of the radiation which is sent towards it).  The radiation power equations for asymptotically AdS black holes at asymptotic frequency are also derived for both the Schwarzschild and 
Reissner-Nordstr\"om solutions which are found same and depend on the  Hawking temperature only. Corresponding to greybody factor  at the low frequency, we demonstrate  the Hawking radiation power equations for asymptotically dS black holes   for both even and odd spacetimes. Here, we find that  the radiation power equation in five-dimensional dS spacetimes
depends on both the horizon radius and Hawking temperature. In case of even spacetime dimensions, we get simpler form of it, however
 it has an infinite sum series of
Hurwitz Zeta function  for the case of odd spacetime dimensions.
It may be noted that  as the greybody factors depend on all frequencies \cite{20}, our formalism can be applied for any given greybody factors. We have plotted diagrams also to understand the behavior of the radiation power equation with respect to both the Hawking temperature and horizon radius.}

This paper is organized as following. In section II, we compute the radiation power equation 
for asymptotically flat spacetimes. Specifically,   we derive radiation power equation for the black holes with the help of greybody factors  at both the low frequency  and  asymptotic frequency for Schwarzschild solution. 
In section III,  we evaluate the  Hawking radiation power equations for the asymptotically AdS black holes. Here, the radiation power equations corresponding to greybody factors at  both low  and  asymptotic  
frequencies  for Schwarzschild solution and Reissner-Nordstr\"om solution are discussed. Further, in section IV, we 
derive   the radiation power equation for greybody factors   at low frequency only for asymptotically dS black holes in even and odd spacetimes.
In the last section, we draw conclusion with final remarks.
\section{Asymptotically Flat Spacetimes}
In this section, we analyse the  the radiation power equation for asymptotically flat black holes. First, we shall derive  the radiation power equation for greybody factor at low frequency   limit  $\omega\ll T_H$, $\omega R_H\ll 1$ for Schwarzschild solution.
Here, the low frequency limit  means that the characteristic scales associated with the
black hole is much smaller than 
the   scalar wave wavelength.
 Then, we discuss the   radiation power equation for greybody factor at 
asymptotic frequency for the case of Schwarzschild solution.
\subsection{Power Equation for Greybody Factors  at Low Frequency}
In this subsection, we compute the radiation power equation for greybody factor at low frequency. Since the greybody factor in the low frequency limit  for asymptotically flat black holes is given by \cite{20}
\bea
\gamma(\omega)=\frac{4\pi\omega^{d-2}R^{d-2}_H}{2^{d-2}[\Gamma(\frac{d-1}{2})]^2},\nn
\eea
where $R_H$ is the horizon radius. Here we clarify that  the above greybody factor is calculated by assuming that  scalar wave, which  spreads from infinity
over whole spacetime,  get reflected by the black hole potential barrier and the transmitted
part of scalar wave, near the  horizon, appears as the incoming radiation to black hole.  We note that,   for low frequency scattering, the greybody factor  is found equal to the absorption probability of the black hole.

For a convenience,  we change dimension  $d\rightarrow (d+1)$\footnote{Throughout the paper $d\rightarrow (d+1)$ represents that $d$ is being replaced by $(d+1)$, where $d$ stands for an arbitrary number of dimensions.} and, hence, the greybody factor for $d+1$ dimensional black hole takes following form:
\bea
\gamma(\omega)=\frac{4\pi\omega^{d-1}R^{d-1}_H}{2^{d-1}[\Gamma(\frac{d}{2})]^2}.\label{gbf-flat}
\eea
 For greybody factor in the low frequency limit as given in
(\ref{gbf-flat}),  the Hawking radiation power equation
for asymptotically flat black holes is given by
\bea
P^{(d+1)}_{low}=\frac{T_H}{2\pi}\int^{\infty}_0 d\omega\frac{4\pi\omega^{d-1}R^{d-1}_H}{2^{d-1}[\Gamma(\frac{d}{2})]^2}\frac{ \omega}{T_H (e^{\frac{\omega}{T_H}}-1)},\nn
\eea
where we have utilized relation (\ref{HRPE}).
After further simplification, this reduces to the following expression:
\bea
P^{(d+1)}_{low}=\frac{  R^{d-1}_H}{2^{d-2}[\Gamma(\frac{d}{2})]^2}\int^{\infty}_0 d\omega\frac{\omega^{d}}{e^{\frac{ \omega}{T_H}}-1}.\label{pr-flat}
\eea
In order to solve above equation, we recall Riemann Zeta function, which is given by,
\bea
\zeta(s)=\frac{1}{\Gamma(s)}\int^{\infty}_0 dy\frac{y^{s-1}}{e^{y}-1} ~~~\text{for}~~ Re(s) > 1.\label{RZF}
\eea
To match this Zeta function with the desired integral, we replace  $y=\frac{ \omega}{T_H}$
in above equation to get,
\bea
\zeta(s)=\frac{1}{T^{s}_{H}\Gamma(s)}\int^{\infty}_0 d\omega\frac{\omega^{s-1}}{e^{\frac{\omega}{T_H}}-1} ~~~\text{for}~~ Re(s) > 1.\label{zf-flat}
\eea
Now, exploiting (\ref{pr-flat}) and (\ref{zf-flat}),   the Hawking radiation power equation
for asymptotically flat black holes is given by
\bea
P^{(d+1)}_{low}&=&C^{(d+1)}_{low} {T^{d+1}_{H}R^{d-1}_{H}},
\eea
where explicit form of $C^{(d+1)}_{low}$ is  $C^{(d+1)}_{low}=\frac{\zeta(d+1)\Gamma(d+1)}{2^{d-2}[\Gamma(\frac{d}{2})]^2}$. Here we see that the Hawking radiation power equation depends on both the Hawking temperature and horizon radius with different power law.
For example, for a black hole in four dimensional spacetime, the   Hawking radiation power equation is proportional to $T_H^4$ and $R_H^2$.
 \begin{figure}[h!]
 \begin{center}$
 \begin{array}{cc}
\includegraphics[width=70 mm]{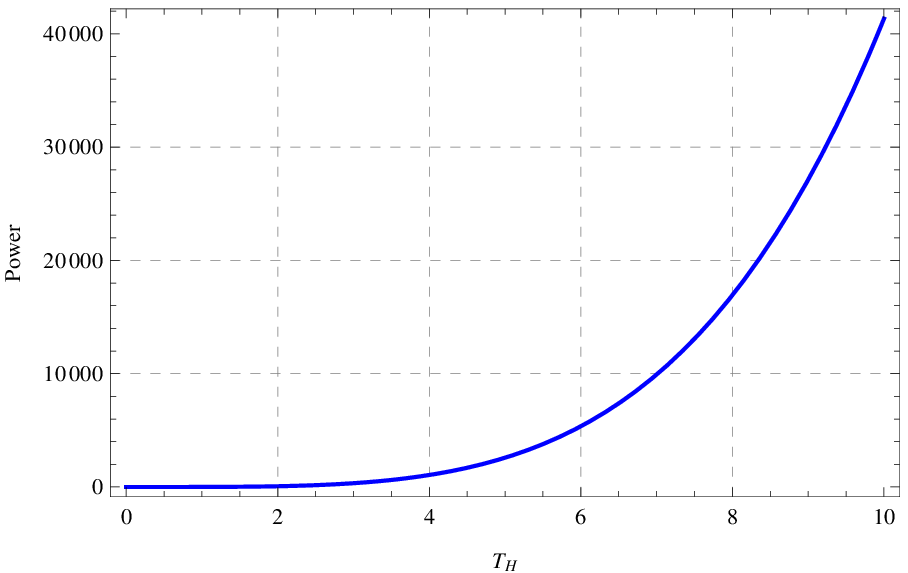}    \ \ \ \ & \includegraphics[width=70 mm]{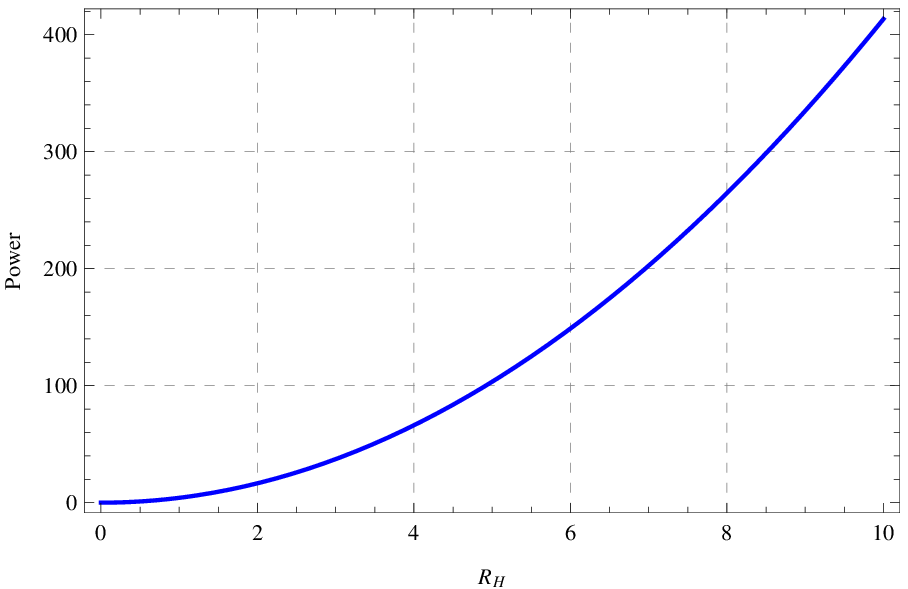} 
 \end{array}$
 \end{center}
\caption{Left: Hawking radiation power versus Hawking temperature for $d=3$ and $R_H=1$.
Right: Hawking radiation power versus horizon radius for $d=3$ and $T_H=1$.}
 \label{fig1}
\end{figure}
The behavior of Hawking radiation power with respect to Hawking temperature and horizon radius
can been seen from plot (\ref{fig1}).
\subsection{Power Equation for Greybody Factors at Asymptotic Frequency }
In this subsection, we determine power equation for greybody factors at asymptotic frequency  
for Schwarzschild solution. 
In order to determine the radiation power equation, we 
first write the Schwarzschild greybody factor 
at asymptotic frequency by  \cite{20},
\bea
\gamma(\omega)=\frac{e^{\frac{ \omega}{T_H}}-1}{e^{\frac{ \omega}{T_H}}+3}.
\eea
The poles of above greybody factor precisely correspond to the asymptotic quasinormal frequencies.
Now, using binomial theorem, we can write the greybody factor 
at asymptotic frequency as
\bea
\le(e^{\frac{ \omega}{T_H}}-1\ri)\le(3+e^{\frac{ \omega}{T_H}}\ri)^{-1}=\sum^{\infty}_{n=1}\frac{(-1)^{n+1}}{3^{n}}\le(e^{\frac{n \omega}{T_H}}-e^{\frac{(n-1) \omega}{T_H}}\ri). \label{gbf-Schz}
\eea
With the help of expression (\ref{HRPE}) and  greybody factor (\ref{gbf-Schz}), the radiation power equation corresponding to the greybody factors at asymptotic frequency for Schwarzschild solution takes the following form:
\bea
P^{(d+1)}_{asym}=\frac{1}{2\pi}\sum^{\infty}_{n=1}\frac{(-1)^{n+1}}{3^{n}}\le(\int^{\infty}_0 d\omega\frac{\omega e^{\frac{n \omega}{T_H}}}{e^{\frac{ \omega}{T_H}}-1}-\int^{\infty}_0 d\omega\frac{\omega e^{\frac{(n-1) \omega}{T_H}}}{e^{\frac{ \omega}{T_H}}-1}\ri).\label{pr-Schz1}
\eea
In order to simplify the above integral, we can use the Hurwitz (generalized Riemann) Zeta function\footnote{Here we note that for this type of Zeta function there are two different forms exists 1) $\zeta(s,q)=\sum ^{\infty}_{a=0}(a+q)^{-s}$ for $q>0$ and 2) $\zeta(s,q)=\sum ^{\infty}_{a=0}[(a+q)^{2}]^{-\frac{s}{2}}$ for $q<0$. We also note that both are identical for $Re(q)>0$.} as the Hurwitz  Zeta function has the following definition:
\bea
\zeta(s,q)=\frac{1}{\Gamma(s)}\int^{\infty}_0 dy\frac{y^{s-1}e^{y(1-q)}}{e^{y}-1} ~~~\text{for}~~Re(s) > 1.\nn
\eea
In order to match the form of above Hurwitz  Zeta function integral with the radiation power equation (\ref{pr-Schz1}), we make  the following identification: $y=\frac{\omega}{T_H}$.
With such identification, the expression of  Hurwitz Zeta function  results to
\bea
\zeta(s,q)=\frac{1}{T^{s}_H\Gamma(s)}\int^{\infty}_0 d\omega\frac{\omega^{s-1}e^{\frac{ \omega}{T_H}(1-q)}}{e^{\frac{ \omega}{T_H}}-1} ~~~\text{for}~~Re(s) > 1.\label{zf-Schz}
\eea
Comparing expressions (\ref{pr-Schz1}) and (\ref{zf-Schz}), one can easily derive the  the radiation power equation corresponding to greybody factors at asymptotic frequency for Schwarzschild solution as
\bea
P^{(d+1)}_{asym}&=&\frac{1}{2\pi}\sum^{\infty}_{n=1}\frac{(-1)^{n+1}}{3^{n}}\le[ {\Gamma(2)\zeta\le(2,1-n\ri)T^{2}_H} - {\Gamma(2)\zeta\le(2,1-(n-1)\ri)T^{2}_H} \ri]\nonumber\\
&=&C^{(d+1)}_{asym} {T^{2}_H},
\eea
where, coefficient $C^{(d+1)}_{asym}=\frac{1}{2\pi}\sum^{\infty}_{n=1}\le[\frac{(-1)^{n+1}}{3^{n}}\le(\zeta\le(2,1-n\ri)-\zeta\le(2,2-n\ri)\ri)\ri]$.
Here, we have used $\Gamma(n)=(n-1)!$ for $\Gamma(2)=1!=1$. 
Remarkably, we find that the 
  radiation power equations corresponding to greybody factors at asymptotic frequency in any arbitrary dimensions are same as it does not depend
on the dimension $d$. Also, we notice that
the radiation power equation corresponding to greybody factors at asymptotic frequency
depends on Hawking temperature only with power law $T^{2}_H$ and does not depend on horizon radius. 
However, for greybody factor at low frequency, it depends on both the Hawking temperature 
and horizon radius. 
\section{Asymptotically AdS Spacetimes}
In this section, we derive   the radiation power equation of black hole in asymptotically  AdS spacetimes corresponding to the greybody factors at both the low frequency and   
asymptotic frequency. 
\subsection{Power Equation for Greybody Factors at Low Frequency}
The greybody factor for asymptotically AdS black
holes, in the low frequency regime  $\hat\omega\ll 1$ is given by   \cite{20},
\bea
\gamma(\hat\omega)=1-\le|\frac{1-z(\hat{\omega})}{1+z(\hat{\omega})}\ri|^{2},\label{gbf-ads}
\eea
where $\hat\omega=\frac{\omega}{k}$ is a dimensionless variable for the frequency and
quantity
\begin{equation}
z(\hat{\omega}) =\frac{\pi}{2^{d-1}[\Gamma(\frac{d}{2})]^2}\frac{\omega^{d-1}}{k^{2d-2}R^{d-1}_H}.\label{zz}
\end{equation} 
Now, we define $\frac{\pi}{2^{d-1}[\Gamma(\frac{d}{2})]^{2} k^{2d-2}R^{d-1}_H}:=\alpha$, so that we can write $z(\hat{\omega}) =\alpha\omega^{d-1}$. Here note that we have used $d\rightarrow (d+1)$ in original form (which can be derived back by using $d\rightarrow (d-1)$ in above expressions). For this greybody factor for asymptotically AdS black
holes  in the low frequency regime (\ref{gbf-ads}), the radiation power equation  (\ref{HRPE})
takes the following form:
\bea
P^{(d+1)}_{low}=\frac{1}{2\pi}\le[\int^{\infty}_0 d\omega\frac{\omega}{e^{\frac{ \omega}{T_H}}-1}-\int^{\infty}_0 d\omega\le(\frac{1-z}{1+z}\ri)^2\frac{\omega}{e^{\frac{ \omega}{T_H}}-1}\ri]\label{pr-ads}.
\eea
Here we see that the first integral of above expression can be solved by using (\ref{zf-flat}), as a result we get $\int^{\infty}_0 d\omega\frac{\omega}{e^{\frac{ \omega}{T_H}}-1}= {\zeta(2)\Gamma(2)T^{2}_H} =\frac{\pi^{2}T^{2}_H}{6 }$. With this simplification, the expression for radiation power   (\ref{pr-ads}) reads,
\bea
P^{(d+1)}_{low}=\frac{\pi}{12} {T^{2}_H} -\frac{1}{2\pi}\int^{\infty}_0 d\omega\le(\frac{1-\alpha\omega^{d-1}}{1+\alpha\omega^{d-1}}\ri)^2\frac{\omega}{e^{\frac{ \omega}{T_H}}-1},\nn
\eea
here we have utilized     $z=\alpha\omega^{d-1}$.
Now, taking $\alpha\omega^{d-1}=\delta$, the above expression can   further be simplified as follows,
\bea
P^{(d+1)}_{low}=\frac{\pi}{12} {T^{2}_H} -\frac{1}{2\pi}\frac{1}{(d-1)\alpha^{\frac{2}{d-1}}}\int^{\infty}_0 d\delta\le(1-\frac{4}{1+\delta}+\frac{4}{(1+\delta)^2}\ri)\frac{\delta^{\frac{3-d}{d-1}}}{\exp\le[\frac{1}{T_H}\le(\frac{\delta}{\alpha}\ri)^{\frac{1}{d-1}}\ri]-1}.\nn
\eea
Here, exploiting binomial expansion for $(1+\delta)^{-1}~\text{and}~(1+\delta)^{-2}$, we find
that $(1+\delta)^{-1}-(1+\delta)^{-2}=\sum^{\infty}_{n=1}(-1)^{n+1}n\delta^{n}$. 
With this result, the above expression for the radiation power equation reduces to,
\bea
P^{(d+1)}_{low}=\frac{\pi}{12} {T^{2}_H} -\frac{1}{2\pi}\frac{1}{(d-1)\alpha^{\frac{2}{d-1}}}\int^{\infty}_0 d\delta\le( 1-4\sum^{\infty}_{n=1}(-1)^{n+1}n\delta^{n}\ri)\frac{\delta^{\frac{3-d}{d-1}}}{\exp\le[\frac{1}{T_H}\le(\frac{\delta}{\alpha}\ri)^{\frac{1}{d-1}}\ri]-1}.\label{pr-ads1}
\eea
In order to solve above integral, 
 we plug $y=\frac{1}{T_H}\le(\frac{\delta}{\alpha}\ri)^{\frac{1}{d-1}}$ in the expression of Zeta function (\ref{RZF}), and we have
\bea
\zeta(s)=\frac{1}{T^{s}_H\Gamma(s)}\frac{1}{\alpha^{\frac{s}{d-1}}(d-1)}\int^{\infty}_0 d\delta\frac{\delta^{\frac{s-d+1}{d-1}}}{\exp\le[\frac{1}{T_H}\le(\frac{\delta}{\alpha}\ri)^{\frac{1}{d-1}}\ri]-1} ~~~\text{for}~~ Re(s) > 1.\label{zf-ads}
\eea
Finally, using (\ref{pr-ads1}) and (\ref{zf-ads}), we get simplified expression for the radiation power equation  corresponding to greybody factor in low frequency limit as 
\bea
P^{(d+1)}_{low}=\sum^{\infty}_{n=1}C_{low}^{(d+1)}\frac{T_{H}^{nd-n+2}}{k^{2n(d-1)} R^{n(d-1)}_H},
\eea
where, $C_{low}^{(d+1)}=\frac{2}{\pi}\le[(-1)^{n+1}\frac{n\pi^n}{2^{n(d-1)\le[\Gamma\le(\frac{d}{2}\ri)\ri]^{2n}}}\zeta(nd-n+2)\Gamma(nd-n+2)\ri]$.
Here, the radiation power equation  has an infinite sum series with terms depending on Hawking temperature  
and horizon radius differently.
 Also, we notice that, contrary to flat spacetime case, the radiation power equation
depends on horizon radius with inverse power law. 

Here, we note that there exists a critical frequency  $\hat{\omega}_c$ for which there is no reflection of scalar wave occurs for black hole which means that the black hole absorbs all of the radiation   sent towards it. Alternatively,
  for emission of scalar wave from the black hole,   all of
the emitted scalar wave at critical frequency should reach the asymptotic region.

In this condition, $z(\hat{\omega}_c)=1$ and critical frequency $\hat{\omega}_c$  simplifies from (\ref{zz}) as \cite{20}
\bea
\hat{\omega}_c=\frac{2\le[\Gamma\le(\frac{d-1}{2}\ri)\ri]^{\frac{2}{d-2}}}{\pi^{\frac{1}{d-2}}}kR_H.\nn
\eea
Here we stress that   the critical frequency can be achieved for  small AdS
black holes  only. 
Now, there are two cases possible in calculation of power radiation equation   for AdS black hole. Firstly,  if we  consider frequencies much lower than the critical frequency ($\hat{\omega}<<\hat{\omega}_c$ )
and secondly, if we   consider  instead frequencies much higher than the critical frequency
($\hat{\omega}>>\hat{\omega}_c$ ).
\subsubsection{Case I:  when $\hat{\omega}<<\hat{\omega}_c$ } 
In this case, the greybody factor for $\hat{\omega}<<\hat{\omega}_c$  is given as follows,
\bea
\gamma(\hat{\omega})=4z(\hat{\omega})=\frac{\pi}{2^{d-2}\le[\Gamma\le(\frac{d-1}{2}\ri)\ri]^{2}}\frac{\hat{\omega}^{d-2}}{\le(kR_H\ri)^{d-2}}.\nn
\eea
Here we notice that the greybody factor is inversely proportional to the area of the black hole, whereas it is proportional to ${\omega}^{d-2}$. Therefore, the frequencies much lower than the critical frequency is identical to $\hat{\omega}<<kR_H$. This means  that
large AdS black holes  (with  $kR_H>>1$)  are always in a frequency regime much lower than the critical frequency.
For convenience (without loss of generality),  we rewrite the above expression of  the greybody factor for $d\rightarrow (d+1)$ as,
 \bea
\gamma(\hat{\omega})=\frac{\pi}{2^{d-1}\le[\Gamma\le(\frac{d}{2}\ri)\ri]^{2}}\frac{\omega^{d-1}}{k^{2d-2}R^{d-1}_H}.\label{gbf-ads1}
\eea
 For this greybody factor, the  power radiation equation  (\ref{HRPE}) is  given by
\bea
P^{(d+1)}_{low-I}=\frac{1}{2^{d}[\Gamma\le(\frac{d}{2}\ri)]^{2}}\frac{1}{k^{2d-2}R^{d-1}_H}\int^{\infty}_0 d\omega\frac{\omega^{d}}{e^{\frac{\omega}{T_H}}-1},\nn
\eea
here, subscript $low-I$ stands for low frequency in region I (i.e. $\hat{\omega}<<\hat{\omega}_c$). In order to obtain explicit expression for  the  power radiation, we need to solve the integral, which can be done easily with the help of Zeta function (\ref{zf-flat}). Hence, we get 
\bea
P^{(d+1)}_{low-I}=C^{(d+1)}_{low-I}\frac{T^{d+1}_H}{k^{2(d-1)}R^{d-1}_H}, \label{ppp}
\eea
where the constant $C^{(d+1)}_{low-I}=\frac{\zeta(d+1)\Gamma(d+1)}{2^{d}\le[\Gamma\le(\frac{d}{2}\ri)\ri]^{2}}$. Here we observe that the radiation power corresponding to greybody factor for frequencies much lower than the critical frequency depends
on both the Hawking temperature with power law $\sim T^{d+1}_H$ and horizon radius 
with power law $\sim R^{d-1}_H$. In four-dimensional spacetimes, the  radiation power equation
reduces to $P^{(d+1)}_{low-I}=\frac{\pi^3}{30}\frac{T_H^4}{k^4R_H^2}$.
\begin{figure}[h!]
 \begin{center}$
 \begin{array}{cc}
\includegraphics[width=70 mm]{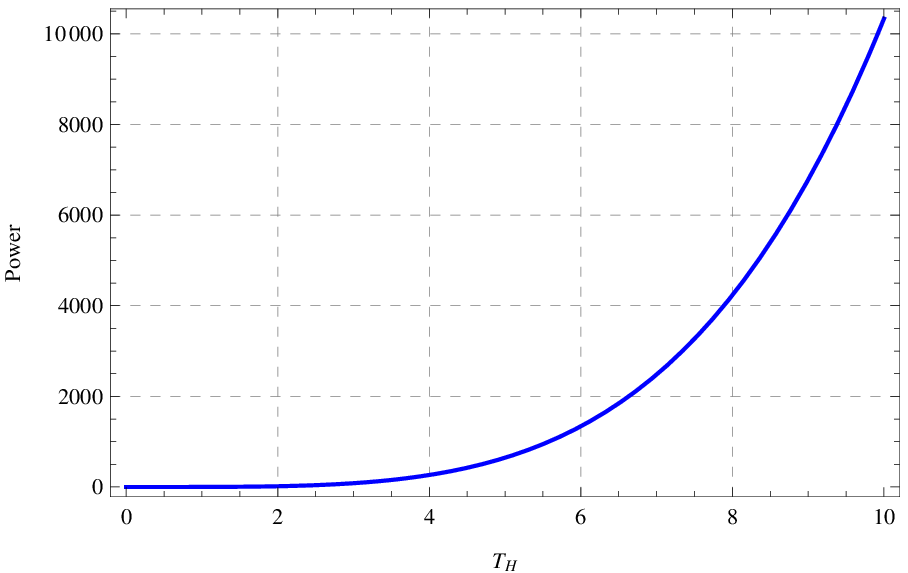}    \ \ \ \ & \includegraphics[width=70 mm]{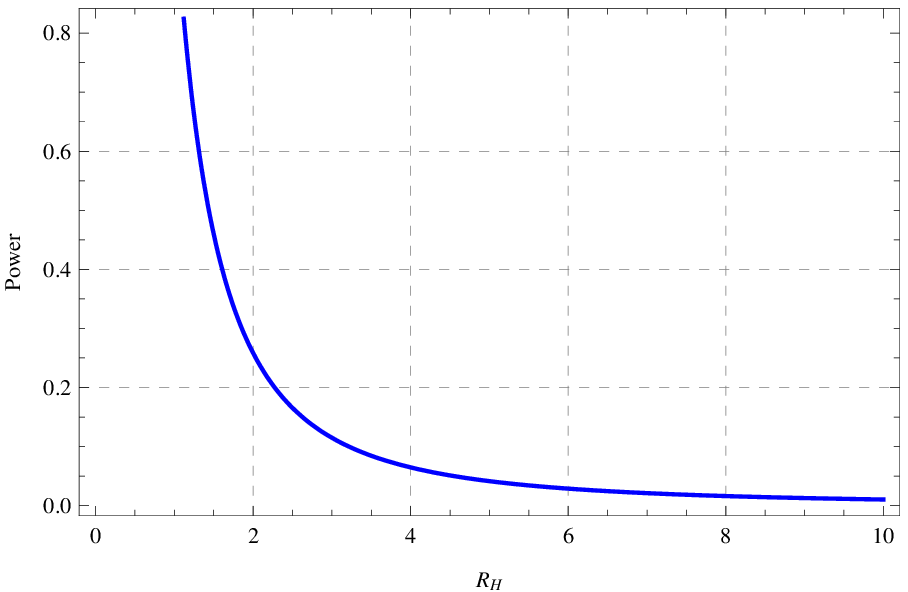} 
 \end{array}$
 \end{center}
\caption{Left: Hawking radiation power versus Hawking temperature for $d=3$ and $k=R_H=1$.
Right: Hawking radiation power versus horizon radius for $d=3$ and $k=T_H=1$.}
 \label{fig2}
\end{figure}
The behavior of Hawking radiation power with respect to Hawking temperature and horizon radius
can be seen in Fig. (\ref{fig2}).
\subsubsection{Case II: when $\hat{\omega}>>\hat{\omega}_c$ } 
In this case of frequencies much higher than the critical frequency, the   greybody factor   for $d\rightarrow (d+1)$ is given by,
\bea
\gamma\le(\hat{\omega}\ri)=\frac{2^{d-1}\le[\Gamma\le(\frac{d}{2}\ri)\ri]^{2}}{\pi}\frac{R^{d-1}_{H}k^{2d-2}}{\omega^{d-1}}.
\eea
Here, greybody factor is proportional to the area of the black hole, whereas it is instead inversely
proportional to $\omega^{d-2}$. Here we note that this frequency regime is possible  for small AdS black holes with $kR_H<<1$ only.
Now, we follow  the same procedure as  discussed in case I and  get the following expression for 
radiation power equation:
\bea
P^{(d+1)}_{low-II}=C^{(d+1)}_{low-II} {T^{-d+3}k^{2(d-1)}R^{d-1}_H},
\eea
where constant $C^{(d+1)}_{low-II}=\frac{2^{d-2}\le[\Gamma\le(\frac{d}{2}\ri)\ri]^{2}}{\pi^{2}}\zeta(-d+3)\Gamma(-d+3)$. Here subscript $low-II$ stands for low frequency in region II (i.e. $\hat{\omega}>>\hat{\omega}_c$). Remarkably, we observe that the radiation power depends both on Hawking temperature with power law $\sim T^{-d+3}$ and horizon radius with power law $\sim R^{d-1}_H$. This clearly means that
for four-dimensional spacetimes the radiation power depends  only on horizon radius  and does not depend on Hawking temperature. However, for space dimensions $d\neq 3$, the  radiation power depends on temperature with different nature. For $d>3$, it depends  on temperature
with inverse power law and, for $d<3$, it depends on temperature with direct power law.
\subsection{Power Equation for Greybody Factors at Asymptotic Frequency}
Remarkably, the greybody factor at asymptotic frequency for  the Schwarzschild solution and the  Reissner-Nordstr\"om solution
is same   and given by \cite{20},
\bea
\gamma(\omega)=1.\label{gbf-ads-Schz}
\eea
For this greybody factor, the Hawking radiation power equation reads,
\bea
P^{(d+1)}_{asym}=\frac{1}{2\pi}\int^{\infty}_0 d\omega\frac{\omega}{e^{\frac{\omega}{T_H}}-1}.\nn
\eea
The above integration can be performed with the help of (\ref{zf-flat}). Thus, we find 
the value of radiation power equation in simplified form as
\bea
P^{(d+1)}_{asym}=C^{d+1}_{asym}{T^{2}_H},\label{pr-ads-Schz}
\eea
where the coefficient $C^{(d+1)}_{asym}=\frac{\zeta(2)\Gamma(2)}{2\pi}=\frac{\pi}{12}$.
Here we conclude that the  radiation power equation corresponding to greybody factor 
at asymptotic frequency depends on Hawking temperature only. 
 
\section{Asymptotically de Sitter Spacetimes}
In this section we shall find the Hawking radiation power equation  corresponding to the greybody factor, at low frequencies, for black holes in asymptotically
dS spacetimes. 
The greybody factor (after considering $d\rightarrow (d+1$)) in this case is given by 
\cite{20}
\bea
\gamma(\omega)=4h(\hat{\omega})(kR_{H})^{(d+1)-2}=4h(\hat{\omega})(kR_{H})^{d-1}\label{gbf-ds-1},
\eea
where  function $h(\hat{\omega})$ for even $(d+1)\geq (3+1)$ is expressed by
\bea
h(\hat{\omega})=\prod_{n=1}^{\frac{(d+1)-2}{2}}\le(1+\frac{\hat{\omega}^2}{(2n-1)^2}\ri)=\prod_{n=1}^{\frac{d-1}{2}}\le(1+\frac{\hat{\omega}^2}{(2n-1)^2}\ri)\label{gbf-ds-2},
\eea
however, for odd $(d+1)\geq (4+1)$ we have,
\bea
h(\hat{\omega})=\frac{\pi\hat{\omega}}{2}\coth\le(\frac{\pi\hat{\omega}}{2}\ri)\prod_{n=1}^{\frac{(d+1)-3}{2}}\le(1+\frac{\hat{\omega}^2}{(2n)^2}\ri)=\frac{\pi\hat{\omega}}{2}\coth\le(\frac{\pi\hat{\omega}}{2}\ri)\prod_{n=1}^{\frac{d-2}{2}}\le(1+\frac{\hat{\omega}^2}{(2n)^2}\ri)\label{gbf-ds-3}.
\eea
Here we see that  $h(\hat{\omega})\rightarrow 1$ and $h(\hat{\omega})\rightarrow 0$ as  $\hat{\omega} \rightarrow 0$, for even and odd spacetime dimension respectively.
Now, we shall calculate the power radiation equation for the given greybody factor by 
 considering specific cases. For example, we  shall study $(a)~(d+1)=4$ and~~(b)~$(d+1)=6$ for the even spacetimes, and $(c)~(d+1)=5$ and~~(d)~$(d+1)=7$ for the  odd spacetimes. 
 
\subsection{Even Spacetimes case I:  (d+1)=4} 
 
In this case, the function $h(\hat{\omega})$ (\ref{gbf-ds-2}) leads to
\bea
h(\hat{\omega})=1+\hat{\omega}^2=1+\frac{\omega^2}{k^2}.\nn
\eea
With this value of $h(\hat{\omega})$, the greybody factor, at low frequencies, for asymptotic 
dS black holes (\ref{gbf-ds-1}) 
reads,
\bea
\gamma(\omega) =4k^2R_{H}^2+4\omega^{2}R_{H}^2\label{gbf-ds-even-1}.
\eea
Now, exploiting relation (\ref{HRPE}), we write the power radiation equation for four-dimensional dS black hole as follows:
\bea
P_{low-even}^{(3+1)}=\frac{2}{\pi}R_{H}^2\le[k^2\int_{0}^{\infty}d\omega\frac{\omega}{e^{\frac{\omega}{T_H}}-1}+\int_{0}^{\infty}d\omega\frac{\omega^{3}}{e^{\frac{\omega}{T_H}}-1}\ri].\label{pr-ds-even-1}
\eea
The integrals of above expression can be solved very easily with the help of 
 Riemann Zeta function (\ref{zf-flat}). After simplification
  the expression (\ref{pr-ds-even-1}) reduces to
\bea
P_{low-even}^{(3+1)}&=&\frac{2}{\pi}R_{H}^2\le(\frac{1}{6} {\pi^{2}k^{2}T_{H}^2}+\frac{6}{90} {\pi^{4}T_{H}^{4}}\ri),\nn\\
&=&\frac{1}{3}\pi {k^2 R_{H}^{2}T_{H}^{2}}+\frac{2}{15}\pi^{3} {R_{H}^{2}T_{H}^{4}}\label{pr-ds-even-2}.
\eea 
It is evident that the radiation power for four-dimensional asymptotically dS black hole 
corresponding to greybody factor at low frequency depends on both the Hawking temperature 
and horizon radius.
\begin{figure}[h!]
 \begin{center}$
 \begin{array}{cc}
\includegraphics[width=70 mm]{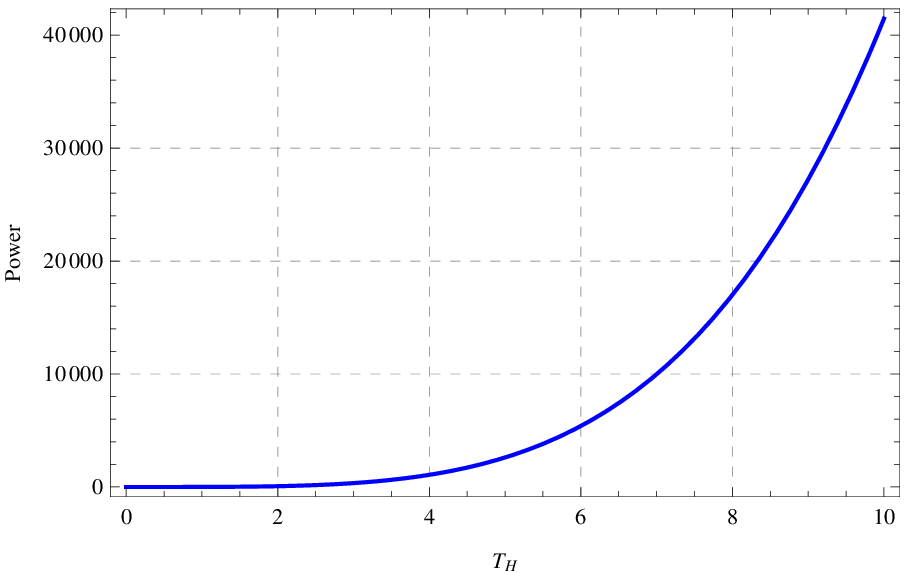}    \ \ \ \ & \includegraphics[width=70 mm]{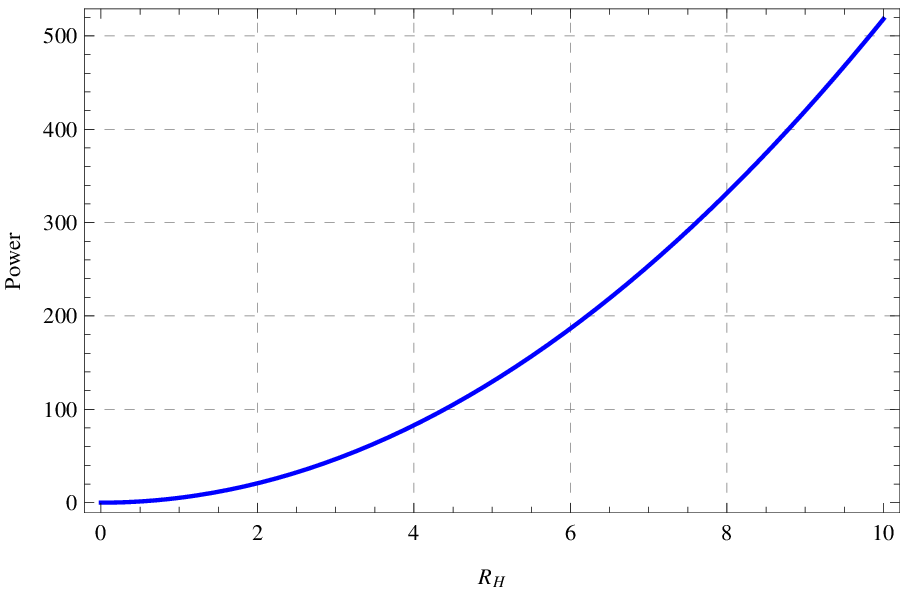} 
 \end{array}$
 \end{center}
\caption{Left: Hawking radiation power versus Hawking temperature for $d=3$ and $k=R_H=1$.
Right: Hawking radiation power versus horizon radius for $d=3$ and $k=T_H=1$.}
 \label{fig3}
\end{figure}
 The behaviors of radiation power for four-dimensional asymptotically dS black hole 
corresponding to greybody factor at low frequency with respect to Hawking temperature 
and horizon radius can be seen in Fig. (\ref{fig3}).
\subsection{Even Spacetimes case II: (d+1)=6}  
\begin{figure}[h!]
 \begin{center}$
 \begin{array}{cc}
\includegraphics[width=70 mm]{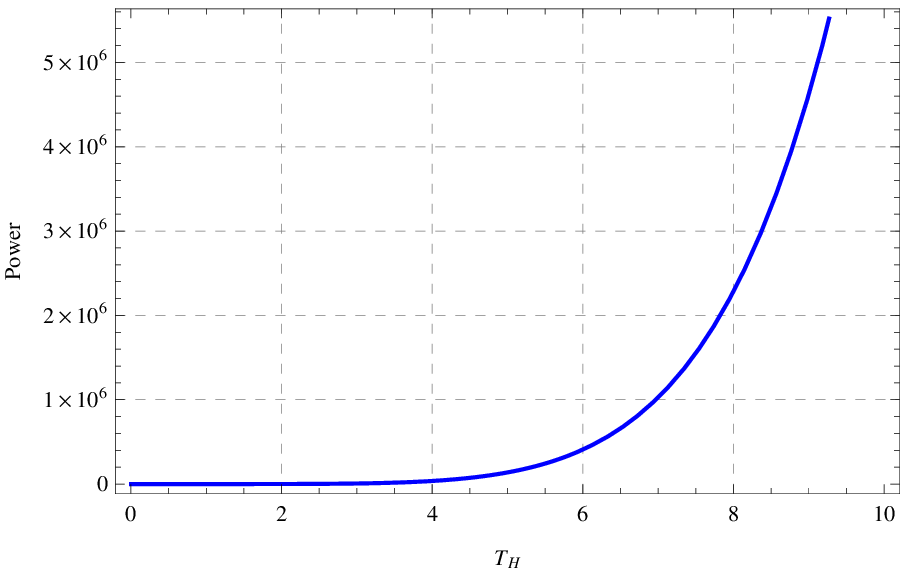}    \ \ \ \ & \includegraphics[width=70 mm]{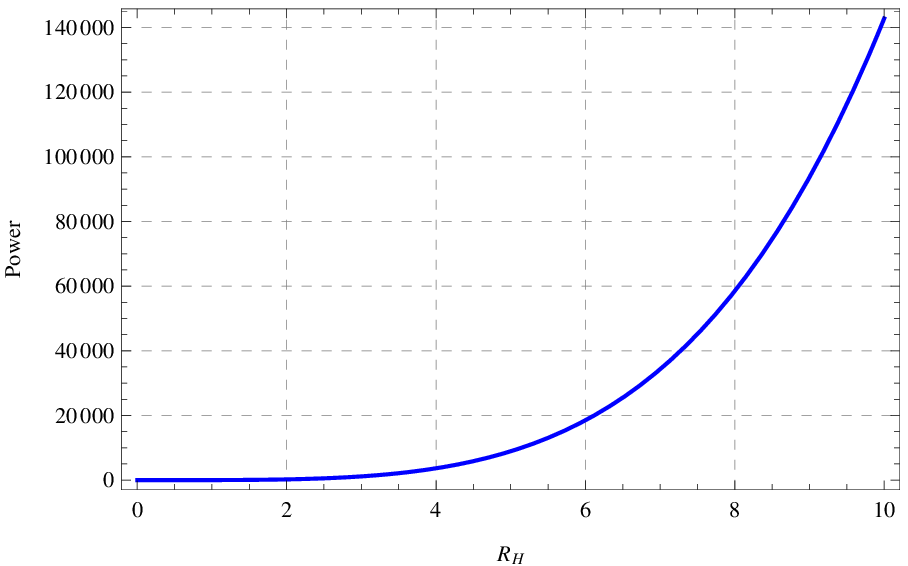} 
 \end{array}$
 \end{center}
\caption{Left: Hawking radiation power versus Hawking temperature for $d=5$ and $k=R_H=1$.
Right: Hawking radiation power versus horizon radius for $d=3$ and $k=T_H=1$.}
 \label{fig4}
\end{figure}

In case  $(d+1)=6$, the expression for $h(\hat{\omega})$ given in  (\ref{gbf-ds-2}) 
has the following form:
\bea
h(\hat{\omega}) =1+\frac{10\omega^{2}}{9k^{2}}+\frac{\omega^{4}}{9k^{4}}.\nn
\eea
With this value of $h(\hat{\omega})$,  the greybody factor, at low frequencies, for asymptotically  dS black holes (\ref{gbf-ds-1}) is given by
\bea
\gamma(\omega) =4k^{4}R_{H}^{4}+\frac{40}{9}\omega^{2}k^{2}R_{H}^{2}+\frac{4}{9}\omega^{4}R_{H}^{4}\label{gbf-ds-even-2}.
\eea
Once the expression for the greybody factor is known, it is matter of calculation to
obtain the  radiation power equation  which utilizes relation (\ref{HRPE}).  
For the greybody factor (\ref{gbf-ds-even-2}), the expression for the radiation power equation  is 
given by
\bea
P_{low-even}^{(5+1)}=\frac{2}{\pi}R_{H}^{4}\le[k^{4}\int_{0}^{\infty}d\omega\frac{\omega}{e^{\frac{\omega}{T_{H}}}-1}+\frac{10}{9}k^{2}\int_{0}^{\infty}d\omega\frac{\omega^{3}}{e^{\frac{\omega}{T_{H}}}-1}+\frac{1}{9}\int_{0}^{\infty}d\omega\frac{\omega^{5}}{e^{\frac{\omega}{T_{H}}}-1}\ri]\label{pr-ds-even-3}.
\eea
In order to solve the integrals in above expression, we utilize the Zeta function
 (\ref{zf-flat}). After doing so, the expression for the radiation power equation  simplified   to
\bea
P_{low-even}^{(5+1)}&=&\frac{2}{\pi}R_{H}^{4}\le[k^{4} {\zeta(2)\Gamma(2)T_{H}^{2}}+\frac{10k^{2}}{9} {\zeta(4)\Gamma(4)T_{H}^{4}}+\frac{1}{9} {\zeta(6)\Gamma(6)T_{H}^{6}} \ri],\nn\\
&=&\frac{1}{3}\pi {k^{4} R_{H}^{4}T_{H}^{2}} +\frac{4}{27}\pi^{3} {k^{2} R_{H}^{4}T_{H}^4}+\frac{16}{567}\pi^{5} {R_{H}^{4}T_{H}^{6}}\label{pr-ds-even-4}.
\eea
 Here, we see that the  the expression for the radiation power equation depends on both the 
 horizon radius and Hawking temperature. In particular, it depends on the sum of different powers of Hawking radiation.  In order to see behavior of  the radiation power equation with respect to 
 horizon radius and Hawking temperature, we plot Fig. (\ref{fig4}).
 
\subsection{Odd Spacetimes case I: (d+1)=5} 
 
The expression for $h(\hat{\omega})$ given in (\ref{gbf-ds-3}) for Odd Spacetimes $(d+1)=5$
takes following value: 
\bea
h(\hat{\omega}) =\frac{\pi\omega}{2k}\coth\le(\frac{\pi\omega}{2k}\ri)\le(1+\frac{\omega^{2}}{4k^{2}}\ri)\label{h-ds-odd-1}.
\eea
Now, in order to simplify above expression we utilize following definition: $\coth(x)=\frac{1+e^{-2x}}{1-e^{-2x}}$. With this definition, we have
\bea
\coth\le(\frac{\pi {\omega}}{2k}\ri) =\le(1+e^{-\pi\frac{\omega}{k}}\ri)\le(1-e^{-\pi\frac{\omega}{k}}\ri)^{-1}.\label{cc}
\eea
With the help of binomial expansion, we can write $\le(1-e^{-\pi\frac{\omega}{k}}\ri)^{-1}=\sum_{m=1}^{\infty}e^{(1-m)\pi\frac{\omega}{k}}$. As a result, the   expression (\ref{cc}) reduces to
 $\coth\le(\frac{\pi\omega}{2k}\ri)=\sum_{m=1}^{\infty}\le(e^{-m\pi\frac{\omega}{k}}+e^{(1-m)\pi\frac{\omega}{k}}\ri).
$
By inserting this value of $\coth\le(\frac{\pi\omega}{2k}\ri)$ into (\ref{h-ds-odd-1}),  
the function $h\le(\frac{\omega}{k}\ri)$ takes the following expression:
\bea
h\le(\frac{\omega}{k}\ri)=\frac{\pi\omega}{2k}\le(1+\frac{\omega^{2}}{4k^{2}}\ri)\sum_{m=1}^{\infty}\le(e^{-m\pi\frac{\omega}{k}}+e^{(1-m)\pi\frac{\omega}{k}}\ri).\nn
\eea
Plugging this value of $h\le(\frac{\omega}{k}\ri)$ in (\ref{gbf-ds-1}),  the greybody factor  for five-dimensional asymptotic dS black
holes at low frequencies  reduces to
\bea
\gamma(\omega)=\frac{2\pi\omega}{k}(kR_{H})^{3}\le(1+\frac{\omega^{2}}{4k^{2}}\ri)\sum_{m=1}^{\infty}\le(e^{-m\pi\frac{\omega}{k}}+e^{(1-m)\pi\frac{\omega}{k}}\ri).\label{gbf-ds-odd-1} 
\eea
Now, it is matter of calculation to evaluate the radiation power equation for a given greybody
factor. So,  utilizing relations (\ref{HRPE}) and (\ref{gbf-ds-odd-1}),  we write the radiation power equation for balck hole in  five-dimensional dS spacetimes  as 
\bea
P_{low-odd}^{(4+1)}&=&\frac{T_{H}}{2\pi}\int_{0}^{\infty}d\omega  \frac{\le(2\pi\omega^2 k^{2}R_{H}^{3}\ri)}{T_{H}\le(e^{\frac{ \omega}{T_{H}}}-1\ri)}  \le(1+\frac{\omega^{2}}{4k^{2}}\ri)\sum_{m=1}^{\infty}\le(e^{-m\pi\frac{\omega}{k}}+e^{(1-m)\pi\frac{\omega}{k}}\ri),\nn\\
&=&  R_{H}^{3}\sum_{m=1}^{\infty}\le[\int_{0}^{\infty}d\omega\le(e^{-m\pi\frac{\omega}{k}}+e^{(1-m)\pi\frac{\omega}{k}}\ri)\le(\frac{k^{2}\omega^{2}}{e^{\frac{ \omega}{T_{H}}}-1}+\frac{1}{4}\frac{\omega^{4}}{e^{\frac{ \omega}{T_{H}}}-1}\ri)\ri]\label{pr-ds-odd-1}.
\eea
In order to simplify the integrals, we 
utilize the Hurwitz Zeta function (\ref{zf-Schz}) and by
doing so,   we get the following explicit expression for the radiation power equation:
\bea
P_{low-odd}^{(4+1)}&=&2k^{2} {R_{H}^{3}T_{H}^{3}} \sum_{m=1}^{\infty}\le[\zeta\le(3,1+\frac{m\pi}{k} {T_{H}}\ri)+\zeta\le(3,1-\frac{(1-m)\pi}{k} {T_{H}}\ri)\ri]\nn\\
&&+6 {R_{H}^{3}T_{H}^{5}}\sum_{m=1}^{\infty}\le[\zeta\le(5,1+\frac{m\pi}{k} {T_{H}}\ri)+\zeta\le(5,1-\frac{(1-m)\pi}{k} {T_{H}} \ri)\ri]\label{pr-ds-odd-2}.
\eea
Here, it is evident that although the  radiation power equation in five-dimensional dS spacetimes depends on both the horizon radius and Hawking temperature but
has an infinite sum series of Hurwitz Zeta function as well.\\
Next, we shall derive the radiation power equation for the $(d+1)=7$. 
 
\subsection{Odd Spacetimes case II: (d+1)=7} 
 
The expression for function $h(\hat{\omega})$   (\ref{gbf-ds-3}) for Odd Spacetimes $(d+1)=7$
is given by:
\bea
h(\hat{\omega}) =\frac{\pi\omega}{2k}\coth\le(\frac{\pi\omega}{2k}\ri)\le(1+\frac{5\omega^{2}}{16k^{2}}+\frac{\omega^{4}}{64k^{4}}\ri).\label{sss}
\eea
Now, plugging the value of $\coth\le(\frac{\pi\omega}{2k}\ri)$ calculated in the above last subsection, the above function $h(\hat{\omega})$ reduces to the following form:
\bea
h\le(\frac{\omega}{k}\ri)=\frac{\pi\omega}{2k}\le(1+\frac{5\omega^{2}}{16k^{2}}+\frac{\omega^{4}}{64k^{4}}\ri)\sum_{m=1}^{\infty}\le(e^{-m\pi\frac{\omega}{k}}+e^{(1-m)\pi\frac{\omega}{k}}\ri)\label{h-ds-odd-2}.
\eea
With this value of  $h\le(\frac{\omega}{k}\ri)$ (\ref{h-ds-odd-2}),   the greybody factor for seven-dimensional asymptotically 
dS black holes at low frequencies  (\ref{gbf-ds-1}) has following value:
\bea
\gamma(\omega)=2\pi\frac{\omega}{k}(kR_{H})^{5}\le(1+\frac{5\omega^{2}}{16k^{2}}+\frac{\omega^{4}}{64k^{4}}\ri)\sum_{m=1}^{\infty}\le(e^{-m\pi\frac{\omega}{k}}+e^{(1-m)\pi\frac{\omega}{k}}\ri)\label{gbf-ds-odd-2}.
\eea
Once the expression for graybody factor of black hole is known, the the Hawking radiation power equation can easily be calculated from relation(\ref{HRPE}).
For a given graybody factor (\ref{gbf-ds-odd-2})at low frequency in $(d+1)=7$ dimensions,   the Hawking radiation power equation  reads,
\bea
P_{low-odd}^{(6+1)}&=&\frac{1}{2\pi}\int_{0}^{\infty}d\omega\frac{\le(2\pi\omega^2 k^{4}R_{H}^{5}\ri)}{ e^{\frac{ \omega}{T_{H}}}-1} \le(1+\frac{5\omega^{2}}{16k^{2}}+\frac{\omega^{4}}{64k^{4}}\ri)  \sum_{m=1}^{\infty}\le(e^{-m\pi\frac{\omega}{k}}+e^{(1-m)\pi\frac{\omega}{k}}\ri),\nn\\
&=&  R_{H}^{5}\sum_{m=1}^{\infty}\le[\int_{0}^{\infty}d\omega\frac{\le(e^{-m\pi\frac{\omega}{k}}+e^{(1-m)\pi\frac{\omega}{k}}\ri)}{e^{\frac{\omega}{T_{H}}}-1}\le( {k^{4}\omega^{2}} +\frac{5}{16} {k^{2}\omega^{4}} +\frac{1}{64} {\omega^{6}} \ri)\ri]\label{pr-ds-odd-3}.
\eea
In order to simplify the above expression, we can use Hurwitz Zeta function 
(\ref{zf-Schz}). By doing so, we get following expression for the Hawking radiation power equation: 
\bea
P_{low-odd}^{(6+1)}&=&2k^{4} {R_{H}^{5}T_{H}^{3}}\sum_{m=1}^{\infty}\le[\zeta\le(3,1+\frac{m\pi}{k} {T_{H}} \ri)+\zeta\le(3,1-\frac{(1-m)\pi}{k} {T_{H}} \ri)\ri]\nn\\
&&+\frac{15k^{2}}{2} {R_{H}^{5}T_{H}^{5}} \sum_{m=1}^{\infty}\le[\zeta\le(5,1+\frac{m\pi}{k} {T_{H}} \ri)+\zeta\le(5,1-\frac{(1-m)\pi}{k} {T_{H}} \ri)\ri]\nn\\
&&+\frac{45}{4} {R_{H}^{5}T_{H}^{7}} \sum_{m=1}^{\infty}\le[\zeta\le(7,1+\frac{m\pi}{k} {T_{H}} \ri)+\zeta\le(7,1-\frac{(1-m)\pi}{k} {T_{H}} \ri)\ri]\label{pr-ds-odd-4}.
\eea
Here, we can see that the  Hawking radiation power equation in  $(d+1)=7$ dimensions also depends on  both the horizon radius and Hawking temperature buth with different power law.
Similar to the previous case, here also,   Hawking radiation power has an infinite sum series of Hurwitz Zeta function.

\section{Conclusions}
 In this paper, we have evaluated the Hawking radiation power equations 
for given greybody factors in asymptotically flat, AdS and dS black holes in $(d+1)$ dimensions.
First of all, we have derived the Hawking radiation power equation for  asymptotically flat  Schwarzschild solution of black hole   corresponding to the greybody factor 
at low frequency. We have solved the  Hawking radiation power equation with the help of Zeta function and found that it depends on both the Hawking temperature and horizon radius with different power law. Moreover, we compute the same for  the greybody factor at  asymptotic frequency  and found that in this region of frequency the Hawking radiation power equation depends on the Hawking temperature only with square power law and remains same for all 
spacetime dimensions. 

 Furthermore, we have evaluated  the 
radiation power equations  for black hole in asymptotically AdS spacetimes.
Here, also we have considered the greybody factors in different regime of frequency
in order to calculate power equations. For the low energy regime, we  note that the critical frequency exists only for small AdS black holes at which  there are no reflection of
radiation for black hole and/or no emission of radiation from the black hole.  Here, we have discussed the cases: 1) if one considers frequencies much lower than the critical frequency  and  2),
if one considers instead frequencies much higher than the critical frequency. 
For former case,  we have observed that the radiation power   depends on both
the Hawking temperature  with power law and horizon radius. However, for later case, remarkably, we observed that though it depends
 on both Hawking temperature   and horizon radius but with different power law. In fact,   for four-dimensional spacetimes the radiation power depends
 on horizon radius only but not on the Hawking temperature.  The radiation power  for greybody factors at asymptotic frequency is independent of horizon radius and dimensionality 
 of spacetime, however,  depends on Hawking temperature only.

 For asymptotically dS spacetime, the Hawking radiation power corresponding to greybody factor
 at low frequency highly depends on dimensionality of spacetime. We have computed this for
 even and odd spacetimes. In case of even dimensions, we have obtained simpler form of
 the radiation power equation which depends on both the  horizon radius and Hawking temperature. 
However, the radiation power equation in odd dS spacetimes although depends on both the horizon radius and Hawking temperature but has an infinite sum series of Hurwitz Zeta function.

It is known that by multiplying the radiation power  expression of black hole with black hole evaporation time, one can estimate the total energy emitted as Hawking radiation  which must be equal to the total mass of the black hole by virtue of 
energy conservation. This might give  an  insight to understand the process of black hole evaporation. Such analysis  a subject of future work.

\end{document}